\documentstyle[12pt,aasms4]{article}
\include{psfig}

\begin{document}
 
\title{Quantitative Morphology of Moderate Redshift Galaxies : How Many
Peculiars Are There ?}

\author{Avi Naim, Kavan U. Ratnatunga AND Richard E. Griffiths}

\affil{The Johns Hopkins University, Department of Physics \& Astronomy, \\
Baltimore, MD 21218, U.S.A.$^1$}

$^1$ current address : Physics Department, Wean Hall, Carnegie Mellon 
University, 5000 Forbes Avenue, Pittsburgh, PA 15213
 
\begin{abstract}

The advent of the Hubble Space Telescope (HST) has provided images of galaxies
at moderate and high redshifts and changed the scope of galaxy morphologies
considerably. It is evident that the Hubble Sequence requires modifications
in order to incorporate all the various morphologies one encounters at 
such redshifts. We investigate and compare different approaches to quantifying 
peculiar galaxy morphologies on images  obtained from the Medium Deep Survey 
(MDS) and other surveys using the Wide Field Planetary Camera 2 (WFPC2) on 
board the HST, in the I band (filter F814W). We define criteria for peculiarity
and put them to use on a sample of 978 galaxies, classifying them by eye as 
either normal or peculiar. Based on our criteria and on concepts borrowed from 
digital image processing we design a set of four purely morphological 
parameters, which comprise the overall texture (or ``blobbiness'') of the 
image; the distortion of isophotes; the filling-factor of isophotes; and the 
skeleta of detected structures. We also examine the parameters suggested by 
Abraham {\it et al.} (1995). An artificial neural network (ANN) is trained to 
distinguish between normal and peculiar galaxies. While the majority of 
peculiar galaxies are disk-dominated, we also find evidence for a significant 
population of bulge-dominated peculiars. Consequently, peculiar galaxies do not
all form a ``natural'' continuation of the Hubble sequence beyond the late 
spirals and the irregulars. The trained neural network is applied to a second, 
larger sample of 1999 WFPC2 images and its probabilistic capabilities are used 
to estimate the frequency of peculiar galaxies at moderate redshifts as $35 
\pm 15 \%$. 

\end{abstract}

\keywords { galaxies: morphology - galaxies: evolution - galaxies: peculiar}

\section{Introduction}

The Hubble Sequence (Hubble 1936) and its revisions (e.g., Sandage 1961;
de Vaucouleurs {\it et al.} 1959; van den Bergh 1976) all address the local 
universe, in effect defining the mainstream of ``normal'' galaxies. It has been
known for a long time that some galaxies do not fit in these schemes. Arp 
(1966) published an atlas of {\it peculiar} galaxies which shows an impressive 
variety of morphologies deemed ``strange'' or ``interesting''. However, 
quantifying the notion of peculiarity is still a challenge. This may be partly 
due to the fact that peculiar galaxies were regarded as rare exceptions 
unrelated to each other, rather than a coherent class (or classes) of galaxies.

The advent of the Hubble Space Telescope (HST) has allowed us to view the
universe of galaxies in much greater depth than ever before. Images in parallel
mode from fields of the Medium Deep Survey (MDS) key project and other 
observations give us a first look at large numbers of galaxies residing at 
moderate redshifts. The galaxy population at a redshift of about $0.5$ looks 
quite different from that of the local universe : A significant population of 
blue irregular galaxies was reported (Griffiths {\it et al.} 1994) which 
appears to account for much of the increase in number counts at faint 
magnitudes (Glazebrook {\it et al.} 1995; Driver {\it et al.} 1995). In 
addition, the incidence of morphological peculiarities among these galaxies 
appears to be higher than in nearby galaxies (e.g., Abraham {\it et al.} 1995, 
hereafter A95). Further out at redshifts exceeding 1 a new morphological class,
called ``chain galaxies'', was recently reported by Cowie {\it et al.} (1995), 
although their nature has already been disputed (Dalcanton \& Shectman 1996).
An inherent problem of such observations is that the rest frame of the 
observed band is shifted bluewards with increasing redshift. It is difficult to
quantify the effect of this shift on the observed images. We discuss below
the effect of observing moderate redshift galaxies in the V band (filter F606W)
as compared with the I band (F814W). Nevertheless, even without accounting for
the shift in rest frame band it appears that the Hubble Sequence needs to be 
extended or replaced by a more general scheme in order to accommodate the 
diversity of shape found among moderate redshift galaxies. The numbers clearly
show that peculiar galaxies can no longer be regarded as rare exceptions.

So far, most published work in this field quoted numbers based on eyeball 
classification of HST images. However, as was pointed out by Lahav \& Naim 
(1996), this approach suffers from two major problems : First, the notion of 
morphological peculiarity is not well defined. There is little agreement - 
even among experts - on what qualifies as a peculiar galaxy. As a result it is 
very difficult to compare results from papers by different authors, and in  
particular a wide range of values are given for the fraction of peculiar
galaxies at moderate redshifts. A second, related problem is the difficulty in
obtaining consistent eyeball classifications for large quantities of galaxies.
A95 realised these problems and and took the important first step of 
introducing a quantitative measure of peculiarity which can be used to 
automatically classify galaxies. 

In this paper we answer two related, yet separate, questions : First, how can
one ``teach'' a computer to tell whether a given galaxy is peculiar or not 
(given an accepted definition for peculiarity). Second, what is the fraction 
of peculiar galaxies at moderate redshifts in the universe. Our premise is that
quantitative measures are crucial if one is to standardise the concept of 
peculiar galaxies. We seek to put this quantitative distinction on firm 
grounds. Since the notion of peculiarity is not well defined we
begin in \S~2 by describing the galaxy samples and discussing our criteria in 
the light of several examples for various types of peculiarity. In \S~3 we 
translate our criteria to four quantitative parameters (hereafter the 
``4P-set'') and show through examples how they are measured. In \S~4 we train 
{\it Artificial Neural Networks} (ANNs) to distinguish between normal and 
peculiar galaxies. We compare the A95 set of parameters (hereafter the ``C-A 
set'') to the 4P set by training three different ANNs. In \S~5 we study the 
limitations of our parameters and examine the ANN success rates as a function 
of magnitude, signal-to-noise ratio and image size, below which our method breaks
down. We then use the trained ANN to classify a larger set of images, 
for which no eyeball classification is available. In \S~6 we examine the 
frequency of peculiarity as a function of the relative importance of the bulge, 
in an attempt to determine whether peculiar galaxies are all disk-dominated. 
The discussion follows in \S~7.

\section{Galaxy Samples and Criteria for Peculiarity}

\subsection{Galaxy Samples}

Our ``sample 1'' consists of some 1059 images in 9 Groth-Westphal strip fields 
(Groth {\it et al.} 1995), and is complete down to a limiting isophotal 
magnitude of $24.0$ in the I band (filter F814W). Of these, 81 are rejected 
either because they are identified by eye as stars or point sources or due to 
low picture quality (e.g., too many bad pixels). The remaining 978 images are 
deemed galaxies. Once the ANN is trained we apply it to the rest of the 
Groth-Westphal strip, down to $I=24.0$. This larger ``sample 2'' contains 2319
images in 18 more fields. One problem with sample 2 is that it contains stars
as well as galaxies, and we need a quantitative criterion for identifying them.
We discuss such a criterion below.

\subsection {A Qualitative Discussion of Criteria for Peculiarity}

A peculiar galaxy is most easily characterised as any galaxy that is not 
morphologically ``normal''. This, in turn, suggests using a representative set
of normal galaxies as templates and comparing any given image to this entire
set. One would then expect the more extreme peculiars to stand out as very 
dissimilar to any of the templates. However, such an approach has already been 
tried without much success by Lahav \& Naim (1996), and they conclude that 
rather than use the whole image as a template one should look for certain 
important {\it features}, those which change the most between normal and 
peculiar galaxies. 

In figure 1 we show twelve examples of I band images, taken from sample 1. 
The top nine of these show galaxies with various peculiar features and the 
bottom three look normal. The peculiar features of the top nine are (going
left to right, then top to bottom) : Very bright ``knots'' on an otherwise 
quiet background; a polar ring galaxy; an apparent merger of two 
galaxies; a weird overall shape; an asymmetric shape; a double core (or an
ongoing merger); a ``knot'' at the end of an edge-on disk; faint spiral arms 
going in two different senses; an arc on one side of the bulge only. Some of
these features are not easy to see in the printed version of the images. From 
an examination of these and other images in sample 1 we came up with two 
characteristics of peculiarity :


\medskip
\begin{enumerate}
\item{The nature of bright, localised structure : Many galaxies are considered 
      peculiar due to the existence of very bright, round features (e.g., 
      ``knots''), as opposed to elongated arms or arm fragments in spirals}.
\item{The degree of symmetry of the image : A different kind of peculiarity 
      is an asymmetric shape. In some cases the asymmetry affects the fainter
      isophotes (e.g., tidal tails), while in others the shape of the image as 
      a whole is rather symmetric, but superposed on it are bright asymmetric
      features}.
\end{enumerate}

\subsection{Eyeball Classifications}

The images were examined in the I (F814W) and V (F606W) bands {\it separately}
by one of us (AN) on a computer screen, and assigned one of the following 
morphological classes : E/S0; Se; Sl; P1; P2. ``Se'' stands for early type
spirals and includes all types in the range S0/a to Sbc. ``Sl'' stands for late
type spirals, and includes all types from Sc through to Im. A distinction 
between two general types of peculiarity is also made : Galaxies with a 
distorted shape are assigned type ``P1'', while those that exhibit localised 
features (e.g., bright ``knots'') are assigned type ``P2''. If a galaxy 
qualifies as peculiar on both counts, the assigned type is ``P2''. The galaxies
in the first three bins are collectively labeled ``normal'' and those in the 
latter two are labeled "peculiar". We would like to point out that there is 
some confusion in the literature between ``Irregular'' and ``Peculiar'' 
galaxies. We define Irregulars as disk dominated galaxies with a small or no 
bulge component, and a fairly flat intensity profile, thus lumping together in 
this class the revised Hubble types Sdm, Sm and Im. Irregulars are therefore
normal galaxies as far as this paper is concerned and belong to the ``Sl'' bin
described above. On the other hand, for a galaxy to be labeled as peculiar at 
least one of the criteria we specify above has to be met. Such galaxies can not
be assigned Hubble types by definition, although they may resemble normal 
galaxies in some respects. 

The approach adopted here is very liberal towards peculiars : even the 
slightest distortions or a single faint knot are enough for a classification 
as a peculiar. This means that galaxies that appear normal but for some small
feature are also called peculiar in our sample. We would like to emphasise that
regardless of the criteria employed, the morphologies of galaxies form 
sequences, and where one draws the line between one type and another is always 
subjective. In recognition of the many border line cases (especially at the 
faint end), each peculiar galaxy is also assigned a certainty index, stating 
simply if its "peculiar" tag is certain or uncertain. The distribution of 
sample 1 galaxies among the classes is shown in table 1 below.

Some $62\%$ of the galaxies receive exactly the same classification, and
the I and V classifications of $80\%$ differ by no more than one type within
each general (normal/peculiar) class. However, there is a ``migration'' 
of galaxies from earlier and normal types in I to later and peculiar types in 
V. This can be seen from the much higher number of galaxies above the diagonal 
in table 1 (297), compared with the number of galaxies below the diagonal (78).
This trend entails a significant increase in the number of peculiars detected 
in V. One possible explanation for this effect is that in the range of 
redshifts covered by our samples (the median redshift of galaxies down to $I 
\sim 22$ is about 0.6) the V filter corresponds to a band that is {\it bluer} 
than the rest-frame B band, while the I filter corresponds to a band slightly 
redder than B in the rest frame. Practically all the existing classification 
schemes for galaxies were defined in the B band for relatively nearby galaxies.
In the local universe the reliability of eyeball classification deteriorates 
fast as one goes bluer than B, and UV morphologies can look very odd compared 
to their B counterparts. Many of the images in our samples are fainter than $I 
\sim 22$, which implies an even higher redshift than 0.6 and consequently an 
even bluer pass band. Ideally, one would wish to have a classification scheme 
for all images in the same rest wavelength, but in practice this is impossible.
On top of this problem, in the Groth-Westphal strip the V band exposures were 
about 30\% shorter than the I band exposures. As a result many images in the V 
band suffer from lower signal-to-noise ratio, which contributes the spurious 
structure and in general misleads the eye to "see" more fragmented structure 
and smaller bulges. Overall, 306 galaxies ($31\%$ of sample 1) are assigned 
peculiar types in the I band and 432 galaxies ($44\%$ of sample 1) are 
assigned peculiar types in the V band.

\section{Quantitative Parameters for Peculiarity}

Following the discussion above, we adopt the I band images for training
the ANN. All of the parameters described below are derived for I band 
images. The images are first treated with a reduction and fitting software 
written by one of us (KUR), which performs a maximum-likelihood analysis of 
the full 2-dimensional image in order to simultaneously fit the sky level, the 
image center, size, position angle and axis ratio, and determine the best 
fitting photometric model (bulge, disk or bulge+disk; see Ratnatunga {\it et 
al., in preparation} for full details). The original images are used for calculating the 
C-A set of A95. In calculating our 4P set we use both the original images and 
the the residual images, left after subtracting the best fitting photometric 
model from the original image.

\subsection{Light Concentration and Asymmetry}

These two parameters are measured following the concepts of Abraham {\it et
al.} 1994) and A95. For the light concentration we use the somewhat different 
definition :

\bigskip
\begin{math}
$$
(1)~~~~~~~~~~~~~~~~~C(\alpha) = \frac {\sum_{r < \alpha \cdot R} I(i,j)}
                                      {\sum_{i,j} I(i,j)}
$$
\end{math}
\bigskip

\noindent where the summation in the denominator is over all image pixels, and
that in the numerator is over all pixels (i,j) whose radius $r$ is less than 
$\alpha$ times the radius of the image, $R$, which is taken to be the length of
the semi-major axis. $r$ is the distance of an image pixel from the image 
center (corrected for ellipticity). $I(i,j)$ is the intensity of pixel (i,j). 
The value $0.3$ is adopted for the parameter $\alpha$ (the same value was used 
by A95, although its definition is slightly different there).

The asymmetry parameter is calculated as in A95 by rotating the image by
$180^{\circ}$ and self-subtracting, and is set equal to the sum of absolute 
values of pixels in the self-subtracted image over the sum of pixels in the
original image :

\bigskip

\begin{math}
$$
(2)~~~~~~~~~~~~~~~~~A = \frac {\sum_{i,j} | I(i,j) - I^{rot}(i,j) |} 
                              {\sum_{i,j} I(i,j)}
$$
\end{math}

\bigskip

\subsection {Measuring Texture}

Many peculiar galaxies appear ``blobby'', while normal galaxies exhibit 
ordered structures or no structure at all. We require that a measure of the
texture take into account the existence of a bright central region in most 
galaxies, and be sensitive to localised structure. The measure we suggest for
HST WFPC2 images is defined as follows : start by binning the pixel intensities
of the raw image into ten linear intensity bins (excluding the lower $5\%$ and 
upper $2\%$ of pixels, to avoid extreme values). Then examine in turn each of 
the pixels whose binned intensity is larger than 5 (the upper half of 
intensities, because it is bright structure that makes the image ``blobby''), 
and denote its distance from the image center (corrected for ellipticity) by 
$d$. Since a typical ``knot'' is no more than 3-4 pixels across, draw a circle 
of radius 0.5 arcsec (5 pixels) around each such pixel, and consider only 
points inside the circle whose (ellipticity-corrected) distance from the image 
center is smaller than $d$. Now calculate what fraction of these points has a 
lower binned intensity than that of the pixel of interest. Figure 2 shows a 
schematic diagram for the process : The solid lines denote intensity contours 
of an idealised elliptical galaxy and the ``X'' denotes the current reference 
pixel. The circle of radius 5 arcsec around it is depicted by the dashed line 
and the open circles are the pixels we examine. In this idealised case it is 
obvious that all of those pixels have intensity equal to or higher than that of
the reference pixel, and our measure would be zero. This measure should be 
close to zero for real ellipticals (all pixels closer to the center are 
necessarily brighter, to within the noise), and should increase as bright 
localised structures become more dominant. Spiral arms are expected to yield 
intermediate values, because some pixels closer to the center will be at least 
as bright (belong to the same arm), while others will be fainter (dust lanes 
by the side of the arm). Repeating this calculation for all pixels whose 
binned intensity is larger than 5 and averaging this measure over all of them 
gives the ``blobbiness'', our first parameter.

\subsection {Measuring Overall Asymmetries}

The asymmetry parameter of A95 is useful, but limited because the detection of 
asymmetric features is a function of their brightness.  This suppresses the
detection of galaxies with faint peculiar features (e.g., tidal tails). As a 
more general approach we define five isophotal levels for the original image, 
separated by equal logarithmic intervals. The faintest $15\%$ of image pixels 
are not considered, because at the faint end pixels form shapes that are very 
sensitive to noise and inaccurate sky subtraction. The pixels of the image are 
then divided into five mutually exclusive isophotal groups, and the geometrical
center of each of these groups is calculated. The distances between the five
centers are worked out and the largest distance, normalised by the image
size, is taken as our second parameter (the isophotal center displacement).

\subsection{Measuring the Distortion of Shapes}

A distortion in the image of a galaxy need not be global. It can happen in
the outer regions, say, as a result of weak interactions with nearby 
galaxies or accretion of gas; or it may only be apparent in the innermost 
regions, say, after a merger has had time to relax but resulted in a double 
core. The galaxy may look different at various isophotal levels, even if its 
isophotes all share roughly the same center. To complement our second parameter,
we define the "maximal elliptical envelope" of each isophote as the smallest 
ellipse containing all the pixels in the isophote. This can be referred to as
the "elliptical hull" of the isophote (in analogy with the "convex hull", e.g.,
Gonzalez and Woods, 1992). The ellipses all share the overall ellipticity and
position angle derived for the image as a whole. The distortion of structure
at a given isophotal level is then defined as the ``filling factor'' of its 
envelope, or more precisely, as the ratio of the number of {\it isophote} 
pixels within the envelope to the total number of pixels in it. This ratio 
approaches $1$ for smooth, axisymmetric galaxies and tends to lower values for 
galaxies with structure. The less ordered the structure, the lower this ratio
gets. We found the filling factor of the third (middle) isophote to be the 
most useful of the five and chose it as our third parameter.

\subsection{Quantifying the Nature of Localised Structure}

Localised structure is best depicted in the residual images, after the 
subtraction of the best-fitting smooth model. The major problem is to tell
residual structure of one kind (spiral arms) from another (e.g., knots). 

In order to quantify the structure we start by binning the intensities of
the residual image into 10 logarithmic bins, which are defined ignoring
the faintest $5\%$ and the brightest $5\%$ of pixels (thus avoiding 
extreme values which could be the result of inaccurate profile subtraction).
We then define as regions of interest all isolated groups of pixels whose
binned intensities are 7 and above, i.e., belonging to the upper $40\%$ of the
intensity scale. We reject regions that are either too small (less than
$1\%$ of the number of image pixels) or too large (more than $30\%$ of the 
image). The former are likely to be just noise patches and the latter are 
unrealistically large structures, which could result from the combined effect
of noise and low surface brightness.

For each region we then work out the "skeleton" (Gonzalez and 
Woods 1992), which can be loosely described as the ``backbone'' of a 
consecutive set of points. We briefly describe the algorithm we used in the
appendix. In figure 3 we demonstrate these ideas with an 
example of four stages in the analysis of an image of a grand-design spiral : 
the original image, the residual (following subtraction of the best fitting 
smooth model), a map of the detected regions with binned pixel intensities 
and a map of their skeleta. We measure the ratio of the number of pixels in 
the skeleton to the number of pixels in the entire region. This ratio is 
closest to zero for perfectly round shapes, where the skeleton shrinks to a 
point, and grows towards 1 with the elongation of the shape. It is superior to 
the axis ratio of a shape (which can be easily obtained from its second moments
matrix), because it can follow winding shapes like spiral arms and uncover 
their true nature. This ratio is then averaged over all regions to get our
fourth parameter.

\subsection{Independence of the Chosen Parameters of each other}

Judging by the definition of our parameters, it may look as if some of them
should convey more or less the same information and could become redundant. We 
check for this possibility by calculating the correlation matrix for our 
parameters (table 2 below). It is clear that there are no strong linear 
correlations between any two of them, although the first two appear more
correlated than the others.

\section{Training an ANN to Classify Normal/Peculiar Galaxies}

\subsection{Training and Testing the ANN}

For an overview of ANNs see Hertz, Krogh and Palmer (1991). Their application 
to morphological classification of galaxies was discussed in detail
elsewhere (Storrie-Lombardi {\it et al.} 1992; Serra-Ricart {\it et al.} 1993; 
Naim {\it et al.} 1995a; Naim 1995b; and especially Lahav {\it et al.}, 1996). 
Briefly, an ANN can be viewed as a non-linear minimising machine which operates
in a multi-dimensional space. Nodes are arranged in layers, where input parameters are 
fed into the nodes of the input layer and the result is read off  the output nodes. 
ANNs which utilise an intermediate (``hidden'') layer between the input and output 
layers are in general more flexible and perform much better than those without it. 
The so-called ``architecture'' of the ANN is the full specification of its nodes and 
their interconnections. The space in which the ANN operates is spanned by the 
connection-strengths (``weights'') between the various nodes. The weights are, in 
effect, the degrees of freedom which are adjusted so as to minimise the error between 
the actual output reading and the desired output. Training the ANN entails repeated 
presentations of patterns from a training set, for which the input parameters 
as well as the desired output values are known. Inputs can be, e.g., 
morphological parameters and output can be, e.g., morphological type. The 
training phase ends when a certain stopping criterion is met. Usually one 
imposes this criterion in terms of the error of the ANN on a different set of 
patterns, known as the testing set. It is not recommended to train and test on the 
same set, since this results in an overfitting of the data and practically 
``memorising'' the particular patterns in that set. If this happens the ANN loses its 
ability to generalise its ``knowledge'' and apply it successfully to patterns 
it has never seen before.

The 978 galaxies in sample 1 are divided into two sets of approximately equal 
sizes, keeping the ratio of normal to peculiar galaxies the same in both sets. 
We use one set as a training set and the other as its testing set. Every ANN is run 
ten times over, each time starting out with a different set of random weights, and its 
resulting classifications are averaged over the ten runs. This is done because in every 
run there is the risk of the ANN getting stuck in a local minimum of its error. Using
different sets of initial weights and running several times over makes it less likely 
to get stuck in the same local minimum time after time and one local minimum will
stand out against the background of the other runs.

The architecture of the ANN depends first of all on the number of parameters
one uses to describe each galaxy. This number determines the number of
input nodes. We have six parameters in all :

\medskip
\begin{itemize}
\item{Light Concentration, following Abraham {\it et al.} 1994, A95}
\item{Asymmetry, following A95.}
\item{Blobbiness}
\item{Isophotal Center Displacement.}
\item{Filling Factor of Middle Isophote.}
\item{Ratio of Skeleton size to Region size.}
\end{itemize}
\medskip

We attempt several combinations of these parameters, as described below. We set the 
number of hidden layer nodes to three and there are two output nodes, one denoting 
normal galaxies and the other peculiars. A typical {\it desired} output vector is 
either (1,0) or (0,1), stating that that galaxy is of the type whose node is set to 1. 
The ANN ends up assigning fractional values to the nodes. These output readings can 
be shown (e.g., Gish 1990; Richard \& Lippman 1991) to approximate the {\it 
Bayesian a-posteriori} probability for each class given the data, in the limit
of very large training sets. The output node with the higher value "wins", and
the corresponding type is assigned to that galaxy. We ran three ANNs, which are 
specified below in terms of their architecture ($N_{input}:N_{hidden}:N_{output}$) 
and the parameters used as inputs.

\medskip
\begin{enumerate}
\item{Using the C-A set of A95 as inputs, architecture 2:3:2}
\item{Using the 4P set, architecture 4:3:2}
\item{Using both sets together, architecture 6:3:2}
\end{enumerate}

Table 3 shows the resulting classification matrices of these runs, where
rows denote eyeball classes and columns denote the resulting classes. Also 
shown are the ``hit rate'' (or completeness) and ``false alarm rate'' (or 
contamination). The hit rate is defined as the fraction of correct 
classifications out of the patterns in each eyeball class. The false alarm 
rate is defined as the fraction of wrong classifications into each eyeball 
class.

The 4P set appears to give a better result than the C-A set, as could be 
expected since it utilises four morphological parameters compared with only
two in the C-A set. In table 4 we present the results of interchanging the 
roles of training and testing sets for all three ANNs. The picture remains 
essentially the same : Using the C-A pair the hit rate over normal galaxies 
decreases a little, while for the 4P set the same effect is compensated by a 
higher hit rate for peculiars. This is also what happens when we use all six
parameters.

\subsection {Improving the Detection of Peculiars}

We concentrate here on the ANNs which utilise the 4P set. Table 5 shows the 
breakdown of successful and unsuccessful classifications of peculiars by the 
ANN, as a function of the certain/uncertain classification index (see \S~2 
above). It is obvious that the misclassifications are much more likely when
the eyeball classification is uncertain. This means that when one applies the 
trained ANN to a set of fresh data (e.g., a catalogue of HST images), it
will identify most clear cases of peculiarity, but only some of 
the less certain cases, as one would expect. It is interesting to note that 
when the ANN is less certain it tends to make more mistakes. Let us define an 
ANN classification as ``uncertain'' if the winning class has a probability of 
less than $0.6$. Then only $12\%$ of the successes of the both ANNs are 
uncertain, while among the misclassifications the fractions rise to $36\%$
for the first ANN and $24\%$ fot the second, twice and more.

Although the above statements are reassuring regarding the overall performance
of the ANN, one may still wish to do a better job of detecting peculiars, even
at a certain price. One way to push the detection level of peculiar galaxies 
up is to change the proportions of normal and peculiar galaxies in the 
training set. As was mentioned above, the ANN approximates Bayesian 
a-posteriori probabilities. These are conditional probabilities for each class 
given the data, which according to Bayes' theorem are proportional to the 
class prior. In other words, if the ANN ``knows'' that there are many more 
normal galaxies than peculiars, it will prefer to guess a normal type whenever 
it is in doubt. If we make the fractions of normals and peculiars equal in the 
training set we remove the class prior and thus give the peculiars a better 
chance. This will come at the price of lowering the success rate for normal 
galaxies and the resulting ANN would not be of use for predicting the overall
morphological mix. In Table 6 we show the results of doing this for both of 
our datasets. The hit rates for peculiars are increased significantly, while 
those of the normal galaxies drop, as expected.

\subsection{Predicting the Morphological Mix}

Apart from classifying individual galaxies correctly, it is interesting to
see if the ANN is capable of predicting the overall frequencies of peculiar 
and normal galaxies in a given dataset. This can be done by utilising 
the ANN probabilities : summing the probabilities assigned to each class
over the entire dataset gives a more accurate and robust picture of the 
morphological mix than counting how many individual galaxies are assigned to 
each class. The reason is that the actual values of the probabilities convey 
much more information than the "all or nothing" approach of choosing a winning 
class and counting whole numbers. In effect, summing up probabilities is not
subject to the ``round-off'' errors that arise when one sums integer numbers.
The picture is not so simple, though : The ANN approximates the conditional 
probability of each class given the data, which depends on the class prior 
probability. As a result, if one presents the ANN with a set of galaxies in 
which the fractions of normals and peculiars are very different from their 
fractions in the training sets, the ANN will not give the correct morphological
mix. The practical implication is that so long as our training sets have {\it 
roughly} the correct mix, we can use the ANN to predict the mix in any large 
dataset. This is the single most important requirement, and a very natural 
one : The training set must represent the ``true'' world faithfully. A 
supervised ANN can therefore be trained to mimic the human decision process
(Naim {\it et al.} 1995a) and thus save much effort and time. The trained ANN
is a parametric, non-linear classifier, and is not limited to a small number of 
parameters. In this respect it is superior to linear discriminatory analysis, and is more
repeatable than eyeball classification.

Even if the training set contains roughly the correct mix of normals and 
peculiars, it is unlikely to give {\it exactly} the same mix. As can be seen
in the top panel of figure 4, when one presents the trained ANN with different 
fractions of peculiars its predicted fraction changes almost linearly with the 
actual fraction, but this linear relation has a slope much flatter than the 
desired value of 1. We produced figure 4 by specifying nine different 
fractions of peculiars, ranging from $20\%$ to $60\%$ in steps of $5\%$. For 
each of these values we select four independent subsets of 40 galaxies each 
from the testing set, and sum the probabilities given by the ANN to get the 
predicted fractions for each of the four subsets. We then average the ANN 
predictions over the four datasets and work out their standard deviations, 
which are depicted by the error bars in the figure. Also shown in the top panel
of figure 4 is the straight line of slope 1 which represents the desired 
relation between the predicted and the actual fractions of peculiars.

The question is then how to correct the mix predicted by the ANN to get a more 
realistic prediction. Since the ANN prediction varies almost linearly with the
true fraction, the simplest thing to do is fit a straight line to the predicted
fractions and use its slope and zero crossing to correct the prediction. The
bottom panel of figure 4 shows the result, and the straight line of slope 1 is 
again plotted just to guide the eye. The error bars are estimated from the 
derived errors in the slope and zero-crossing of the fitted line. This 
procedure gives us an estimate of the error in predicting the morphological 
mix, and can be used for any set of galaxies with a peculiar fraction between
$20\%$ and $60\%$.

\section{Limits of Applicability}

We now turn to study the accuracy of ANN classifications as a function of 
various limiting quantities. In figure 5 we show three histograms describing 
the hit rates of the ANN as a function of magnitude, signal-to-noise ratio and 
image size. Magnitudes are isophotal (Ratnatunga {\it et al., in preparation}).
Signal-to-noise ratios are integrated over all pixels associated with the
image (as determined by the object detection algorithm), for which the 
individual ratio is larger than 1. This limit is imposed in order to avoid 
inclusion of large patches of sky in the image. Image size is the number of 
pixels whose intensity is at least $5~\sigma$ above sky, i.e., the bright 
pixels of the image. The success rate for normal galaxies drops almost steadily
as one goes fainter and as the ratio of signal-to-noise decreases. A similar 
trend exists when the number of bright pixels decreases. However, the picture 
is less clear with respect to the peculiar galaxies : The trend with magnitude 
is "bumpy" as is the trend with signal-to-noise, though the dependence on the 
number of bright pixels appears more stable. We conclude that our analysis does
not break down sharply at any point, although if one's concern is mainly with 
the identification of peculiars we would advise keeping the signal-to-noise 
ratio above 180 and the number of bright pixels above 50.

\section{The Distribution of Bulge-to-Disk Ratios}

We now turn to apply the trained ANN to the larger sample 2, which was not 
classified by eye. This sample contains 2319 images and includes stars as well 
as galaxies. A quantitative criterion that will allow us to reject stars 
without resorting to eyeball classification is required. We use the maximum 
likelihood software (Ratnatunga {\it et al., in preparation}) to fit a smooth 
photometric model to each galaxy image. The model has 10 degrees of freedom, 
among which is the log of the effective half light radius. This parameter is
expected to be very small for stars and larger for galaxies. We test this
expectation in figure 6, where we show two histograms depicting the 
distribution of half light radii in the eyeballed sample 1. The dashed line 
depicts images which were eyeballed as either stars or point sources, and the 
solid line depicts those that were deemed galaxies. There is a sharp 
distinction between the two populations at a half light radius of roughly 0.1 
arcsec, which corresponds to a single pixel in WFPC2 images. This value is a 
function of magnitude, and compact, high redshift galaxies may be excluded. 
However, we do not have any other means for removing stars from our sample,
and regardless of the true nature of an object, if its half light radius is
less than one pixel it contains very little morphological information for our 
analysis. We therefore reject from sample 2 those images whose best fitted half
light radius is less than this value. The number of rejected images is 320, 
and we are left with 1999 galaxies in this sample. These are then classified 
using the trained ANN.

One of the parameters fitted by the maximum likelihood algorithm is the 
bulge-to-total ratio, which is defined as the ratio of light in the fitted
bulge component to the sum of the light in the fitted bulge and disk 
components (integrated to infinity for both). This B/T ratio is a quantitative 
measure of the dominance of the bulge in the underlying photometric model, and 
by construction is less susceptible to bright overlying features than is the 
cruder light concentration. The software makes a decision, based on the 
signal-to-noise ratio of the image, whether to fit a full bulge+disk model or
to fit only a pure bulge or a pure disk. In these latter cases the B/T ratio is
automatically fixed at 1 or 0 (respectively). In figure 7 we plot the Light
Concentration index against the B/T ratio, and the large numbers of points at
B/T values of 0 and 1 are a testimony of the significant fractions of images
that were fitted as either pure bulges or pure disks. The two quantities are
not very correlated, which is surprising. Since the B/T has been checked for
simulated images and gave very good results (Ratnatunga {\it et al., in 
preparation}), we conclude that the light concentration index is too sensitive 
to structure superimposed on the underlying smooth model. We therefore 
concentrate in what follows on the B/T ratio as the primary indicator of bulge 
dominance. In figure 8 we show the distribution of the B/T ratios among normal 
and peculiar galaxies, as classified by the ANN. The plot includes galaxies for 
which the estimated error in the B/T ratio was less than $0.1$ (the width of a 
single bin in the histograms). This excludes 122 out of the 1999 galaxies. It 
is clear that the fraction of disk dominated peculiars is larger than the 
corresponding fraction for normal galaxies. Nevertheless, about $17\%$ of all 
peculiars have a B/T ratio larger than $0.5$, which we take as ``bulge dominated''. 
While this is less than the corresponding $23\%$ among normals, it still implies a 
significant population.

\section{Discussion}

The frequencies of various morphological types that are quoted in the 
literature are almost invariably estimated on the basis of eyeball 
classifications. As a result it is difficult to compare work done by different
observers because they have different criteria for peculiarity. A95 made the
crucial step of introducing a {\it quantitative} criterion, but as shown 
above it gives only a low hit rate for peculiars, together with a significant 
contamination by normal galaxies. The confusion between normal and peculiar 
galaxies can be seen in figures 5 and 6 of their paper, where one also sees 
several peculiars lying in the region dominated by ellipticals. Their use of 
the light-concentration index as a second parameter precludes bulge dominated 
peculiar galaxies from the outset. It might be better to define peculiarity on 
the basis of shape alone, without any reference to the distribution of light. 

We therefore adopt a purely morphological approach, concentrating on the 
features that make a galaxy appear peculiar to the eye. This then gives us the 
freedom to define the population of peculiars on a morphological basis and 
examine the distribution of bulge-to-disk ratios in this population separately.
As we show above, there is indeed a population of bulge-dominated peculiars.
This population probably corresponds to the ``blue nucleated galaxies'' (BNGs)
found in the Canada-France Redshift Survey (CFRS, Schade {\it et al.} 1995; 
Schade {\it et al.} 1996), although this correspondence is still to be 
confirmed. Unlike CFRS we do not have redshifts for the galaxies in our sample
but our imaging is much better, coming from HST. The success of fitting a bulge
photometric model depends strongly on the point spread function, and therefore
it is not obvious that ground based images deemed bulge dominated will maintain
this quality when imaged by HST (see Ratnatunga {\it et al., in preparation} for
a fuller discussion of this point).

We apply the ANNs trained using the 4P set to a collection of 1999
galaxies taken from the other 18 fields of the Groth-Westphal strip (Groth {\it
et al.} 1995). Summing up the probabilities for peculiars over this entire 
set and inverting to get the true frequency of peculiars at moderate redshifts,
we find this frequency to be $35 \pm 15 \%$. The error we obtain is the first 
to date to rely on statistical estimates rather than on the much cruder 
estimate of eyeball classification errors for individual galaxies. The fact 
that the error bar is large is a reflection of the uncertainties involved in 
the entire process.

It is quite clear now that so long as the automated procedure relies on 
eyeball classifications it will be partly subjective and difficult to 
standardise. What is called for at this stage is a set of quantitative 
parameters that adequately describe the appearance of galaxies. In this paper 
we provide such a set. Once the parameter space is fixed, one way to proceed is
to look for correlations with other sources of information (e.g., colours, 
luminosities), and examine whether in this space the locus of galaxies of, 
e.g., a certain colour is significantly different that the locus of galaxies 
of a different colour. 

Morphological classification is a good starting point for studying galaxy 
evolution, by virtue of the large numbers of available images. However, it can 
only serve as a first step, because galaxy shapes form a continuum rather than
break down to separate groups. In addition, the classes one defines by eye 
might not always have a one-to-one correlation with the dynamics and chemistry 
of galaxies. In this paper we use morphology as tool for singling out 
interesting galaxies for more detailed studies. The next logical step is to 
examine where different populations of galaxies reside in the space of our 
morphological parameters, without recourse to any kind of classification.
Work along these lines is currently in progress (Naim {\it et al., in 
preparation}). Hopefully, once large 
quantities of physical data (e.g., spectra, rotation curves) are available for
these galaxies, we will be able to quantify and understand galaxy evolution 
much better.

\newpage

\bigskip
{\bf ACKNOWLEDGEMENTS}
\bigskip

We would like to thank Ofer Lahav for statistical insight and Brian Ripley for
allowing us to use his ANN code. Stefano Casertano and Myungshin Im read the 
manuscript and raised valuable points and Eric Ostrander's contribution to 
the MDS pipeline processing of the images was invaluable. The referee made useful 
comments and we thank him for helping us make this paper more readable. This research was
supported by funding from the HST Medium Deep Survey under GO grants p2684 {\it et seq.}

\bigskip

{\bf APPENDIX }: Algorithm for Calculating the Skeleton of an Image

\medskip

The following algorithm appears in Gonzalez \& Woods (1992) and was adopted by 
us for calculating the skeleta of galaxy features. This algorithm is suitable 
for skeletoning binary regions, i.e., regions in which each pixel is either 
part of the feature whose skeleton we seek (value 1) or does not belong to it 
at all (value 0). No reference is made to the actual intensity of the pixels. 

Denote a pixel in the image by $p1$ and its nearest eight neighbours by $p2,
\ldots,p9$, starting with the one directly above it and going clockwise 
around it (figure 9). Define as contour points pixels whose value is 1, with at
least one neighbour whose value is zero. The skeletoning process is iterative 
and consists of repeating a set of two steps. In each step we examine value 1 
pixels and those that comply with a given set of criteria are flagged for 
deletion. The values of pixels flagged for deletion are all changed to 0 at the
end of each step. The process stops once no more pixels are flagged for 
deletion. 

In step 1 of each iteration a pixel $p1$ is flagged for deletion if it fulfills
all of the following requirements :

\begin{itemize}
\item{the number of value 1 neighbours is in the range $[2,6]$}
\item{the number of changes from value 0 to value 1, as one goes clockwise 
around $p1$, is 1.}
\item{the value of at least one of $p2, p4$ and $p6$ is zero.}
\item{the value of at least one of $p4, p6$ and $p8$ is zero.}
\end{itemize}

In step 2 the first two requirements remain unchanged, but the last two are 
modified as follows :

\begin{itemize}
\item{the value of at least one of $p2, p4$ and $p8$ is zero.}
\item{the value of at least one of $p2, p6$ and $p8$ is zero.}
\end{itemize}

This procedure is somewhat dependent on the order of the steps and the 
resolution of the image (a small number of pixels will normally imply a highly 
discretised contour and consequently a less dependable skeleton).

\newpage

\begin{table}
\centering
\caption{Breakdown of Morphological Eyeball Classifications in I and V bands.}
\halign{\hfil#\hfil &\quad\hfil#\hfil &\quad\hfil#\hfil &\quad\hfil#\hfil 
 &\quad\hfil#\hfil &\quad\hfil#\hfil &\quad\hfil#\hfil &\quad\hfil#\hfil 
&\quad\hfil#\hfil \cr
 {\bf I / V} & {\bf E/S0} & {\bf Se} & {\bf Sl} & & {\bf P1} & {\bf P2} & & 
 {\bf Total} \cr
{\bf E/S0}   & 46 & 21 &  5 & &  0 &  0 & & 72 \cr
{\bf Se}     & 12 &158 & 80 & & 37 & 28 & &315 \cr
{\bf Sl}     &  0 & 17 &174 & & 46 & 48 & &285 \cr
\cr
{\bf P1}     &  0 &  9 & 10 & & 91 & 32 & &142 \cr
{\bf P2}     &  0 &  1 & 13 & & 16 &134 & &164 \cr
\cr
{\bf Total}  & 58 &206 &282 & &190 &242 & &    \cr}

\label{eyeball}
\end{table}

\begin{table}
\centering
\caption{Correlation Matrix for the Parameters of the 4P Set.}
\halign{\hfil#\hfil &\quad\hfil#\hfil &\quad\hfil#\hfil &\quad\hfil#\hfil 
 &\quad\hfil#\hfil &\quad\hfil#\hfil \cr
  &  {\bf Blobb-} & {\bf Iso.}  & {\bf Iso. } & {\bf Skel.} \cr
  &  {\bf iness}  & {\bf Disp.} & {\bf Fill.} & {\bf Ratio} \cr
{\bf Blobbiness}   &  1.00 & -0.47 &  0.14 & -0.21 \cr
{\bf Iso. Disp.}   & -0.47 &  1.00 &  0.24 & -0.02 \cr
{\bf Iso. Fill.}   &  0.14 &  0.24 &  1.00 & -0.22 \cr
{\bf Skel. Ratio}  & -0.21 & -0.02 & -0.22 &  1.00 \cr}
\label{corr_mat}
\end{table}

\begin{table}
\centering
\caption{Results of ANN trained on set 1 and tested on set 2.}
\halign{\quad\hfil#\hfil \cr
{\bf Using the C-A Pair (Light Concentration and Asymmetry)} \cr
\cr}
\halign{\hfil#\hfil &\quad\hfil#\hfil &\quad\hfil#\hfil &\quad\hfil#\hfil 
 &\quad\hfil#\hfil \cr
 & {\bf ANN} & {\bf ANN} & {\bf Hit} & {\bf False} \cr 
 & {\bf Normal} & {\bf Peculiar} & {\bf Rate} & {\bf Alarm} \cr 
{\bf Eyeball Normal} & 316 & 19 & 94\% & 28 \% \cr
{\bf Eyeball Peculiar} & 120 & 34 & 22\% & 36 \% \cr
\cr
\cr}
\halign{\quad\hfil#\hfil \cr
{\bf Using the 4P Set} \cr
\cr}
\halign{\hfil#\hfil &\quad\hfil#\hfil &\quad\hfil#\hfil &\quad\hfil#\hfil 
 &\quad\hfil#\hfil \cr
 & {\bf ANN} & {\bf ANN} & {\bf Hit} & {\bf False} \cr 
 & {\bf Normal} & {\bf Peculiar} & {\bf Rate} & {\bf Alarm} \cr 
{\bf Eyeball Normal} & 291 & 44 & 87\% & 23 \% \cr
{\bf Eyeball Peculiar} & 87 & 67 & 44\% & 40 \% \cr
\cr
\cr}

\halign{\quad\hfil#\hfil \cr
{\bf Using All Six Parameters} \cr
\cr}
\halign{\hfil#\hfil &\quad\hfil#\hfil &\quad\hfil#\hfil &\quad\hfil#\hfil 
 &\quad\hfil#\hfil \cr
 & {\bf ANN} & {\bf ANN} & {\bf Hit} & {\bf False} \cr 
 & {\bf Normal} & {\bf Peculiar} & {\bf Rate} & {\bf Alarm} \cr 
{\bf Eyeball Normal} & 299 & 36 & 89\% & 22 \% \cr
{\bf Eyeball Peculiar} & 83 & 71 & 46\% & 34 \% \cr}
\label{ann1}
\end{table}

\begin{table}
\centering
\caption{Results of ANN trained on set 2 and tested on set 1.}

\halign{\quad\hfil#\hfil \cr
{\bf Using the C-A Pair} \cr
\cr}
\halign{\hfil#\hfil &\quad\hfil#\hfil &\quad\hfil#\hfil &\quad\hfil#\hfil 
 &\quad\hfil#\hfil \cr
 & {\bf ANN} & {\bf ANN} & {\bf Hit} & {\bf False} \cr 
 & {\bf Normal} & {\bf Peculiar} & {\bf Rate} & {\bf Alarm} \cr 
{\bf Eyeball Normal} & 307 & 28 & 92\% & 28 \% \cr
{\bf Eyeball Peculiar} & 117 & 36 & 24\% & 44 \% \cr
\cr
\cr}

\halign{\quad\hfil#\hfil \cr
{\bf Using the 4P Set} \cr
\cr}
\halign{\hfil#\hfil &\quad\hfil#\hfil &\quad\hfil#\hfil &\quad\hfil#\hfil 
 &\quad\hfil#\hfil \cr
 & {\bf ANN} & {\bf ANN} & {\bf Hit} & {\bf False} \cr 
 & {\bf Normal} & {\bf Peculiar} & {\bf Rate} & {\bf Alarm} \cr 
{\bf Eyeball Normal} & 305 & 30 & 91\% & 22 \% \cr
{\bf Eyeball Peculiar} & 86 & 67 & 44\% & 31 \% \cr
\cr
\cr}

\halign{\quad\hfil#\hfil \cr
{\bf Using All Six Parameters} \cr
\cr}
\halign{\hfil#\hfil &\quad\hfil#\hfil &\quad\hfil#\hfil &\quad\hfil#\hfil 
 &\quad\hfil#\hfil \cr
 & {\bf ANN} & {\bf ANN} & {\bf Hit} & {\bf False} \cr 
 & {\bf Normal} & {\bf Peculiar} & {\bf Rate} & {\bf Alarm} \cr 
{\bf Eyeball Normal} & 304 & 31 & 91\% &21 \% \cr
{\bf Eyeball Peculiar} & 80 & 73 & 48\% & 30 \% \cr
\cr
\cr}
\label{ann2}
\end{table}

\begin{table}
\centering
\caption{Breakdown of ANN Successes and Failures in classifying Peculiars,
as a function of the degree of Certainty in Eyeball Classifications.}

\halign{\quad\hfil#\hfil
{\bf For the ANN trained on set 1.} \cr
\cr}
\halign{\hfil#\hfil &\quad\hfil#\hfil  &\quad\hfil#\hfil \cr
 & {\bf Eyeball} & {\bf Eyeball} \cr
 & {\bf Certain} & {\bf Uncertain} \cr
{\bf ANN Suceesses} & 48 & 19 \cr
{\bf ANN Failures} & 40 & 47 \cr
\cr
\cr}

\halign{\quad\hfil#\hfil
{\bf For the ANN trained on set 2.} \cr
\cr}
\halign{\hfil#\hfil &\quad\hfil#\hfil &\quad\hfil#\hfil \cr
 & {\bf Eyeball} & {\bf Eyeball} \cr
 & {\bf Certain} & {\bf Uncertain} \cr
{\bf ANN Suceesses} & 52 & 15 \cr
{\bf ANN Failures} & 35 & 51 \cr}
\label{ann_sf}
\end{table}

\begin{table}
\centering
\caption{Results of ANN Trained on Equal Numbers of Normals and Peculiars. 
Compare with tables 3 and 4.}

\halign{\quad\hfil#\hfil \cr
{\bf Training on Set 1} \cr}
\halign{\hfil#\hfil &\quad\hfil#\hfil &\quad\hfil#\hfil &\quad\hfil#\hfil 
 &\quad\hfil#\hfil \cr
 & {\bf ANN} & {\bf ANN} & {\bf Hit} & {\bf False} \cr 
 & {\bf Normal} & {\bf Peculiar} & {\bf Rate} & {\bf Alarm} \cr 
{\bf Eyeball Normal} & 108 & 46 & 70\% & 30 \% \cr
{\bf Eyeball Peculiar} & 46 & 108 & 70\% & 30 \% \cr
\cr
\cr}

\halign{\quad\hfil#\hfil \cr
{\bf Training on Set 2} \cr}
\halign{\hfil#\hfil &\quad\hfil#\hfil &\quad\hfil#\hfil &\quad\hfil#\hfil 
 &\quad\hfil#\hfil \cr
 & {\bf ANN} & {\bf ANN} & {\bf Hit} & {\bf False} \cr 
 & {\bf Normal} & {\bf Peculiar} & {\bf Rate} & {\bf Alarm} \cr 
{\bf Eyeball Normal} &   95 & 59 & 62\% & 27 \% \cr
{\bf Eyeball Peculiar} & 35 & 118 & 77\% & 33 \% \cr
\cr
\cr}
\end{table}

\newpage

\bigskip
{\bf Figure Captions :}
\bigskip

Figure 1 : twelve Galaxies from our Sample 1. The top Nine were Classified as
Peculiars and the Bottom Three as Normal Galaxies.

Figure 2 : derivation of the Blobbiness Parameter : Idealised Elliptical Case.

Figure 3 : skeletoning of a Grand-Design Spiral. Top Left : The Original Image;
Top Right : The Residual Image; Bottom Left : Detected Regions; Bottom Right :
Skeleta of the Regions.

Figure 4 : ANN Prediction of the Morphological Mix vs. the True Mix. Top : 
Before applying correction; Bottom : After applying correction.

Figure 5 : the Dependence of ANN Success Rates on Magnitude (Top Left), 
Integrated Signal-to-Noise Ratio (Top Right), and Image Size (Bottom Right). 
Normal galaxies are depicted by the solid line, Peculiars by the dashed line.

Figure 6 : the Distributions of Fitted Half Light Radii for Eyeball Point 
Sources and Eyeball Galaxies.

Figure 7 : the Lack of Correlation between Light Concentration and the Maximum
Likelihood B/T Ratio.

Figure 8 : the Distributions of B/T Ratios among Normal and Peculiar Galaxies.

Figure 9 : the notation used for the neighbours of a given pixel $p1$ in the
skeletoning algorithm.

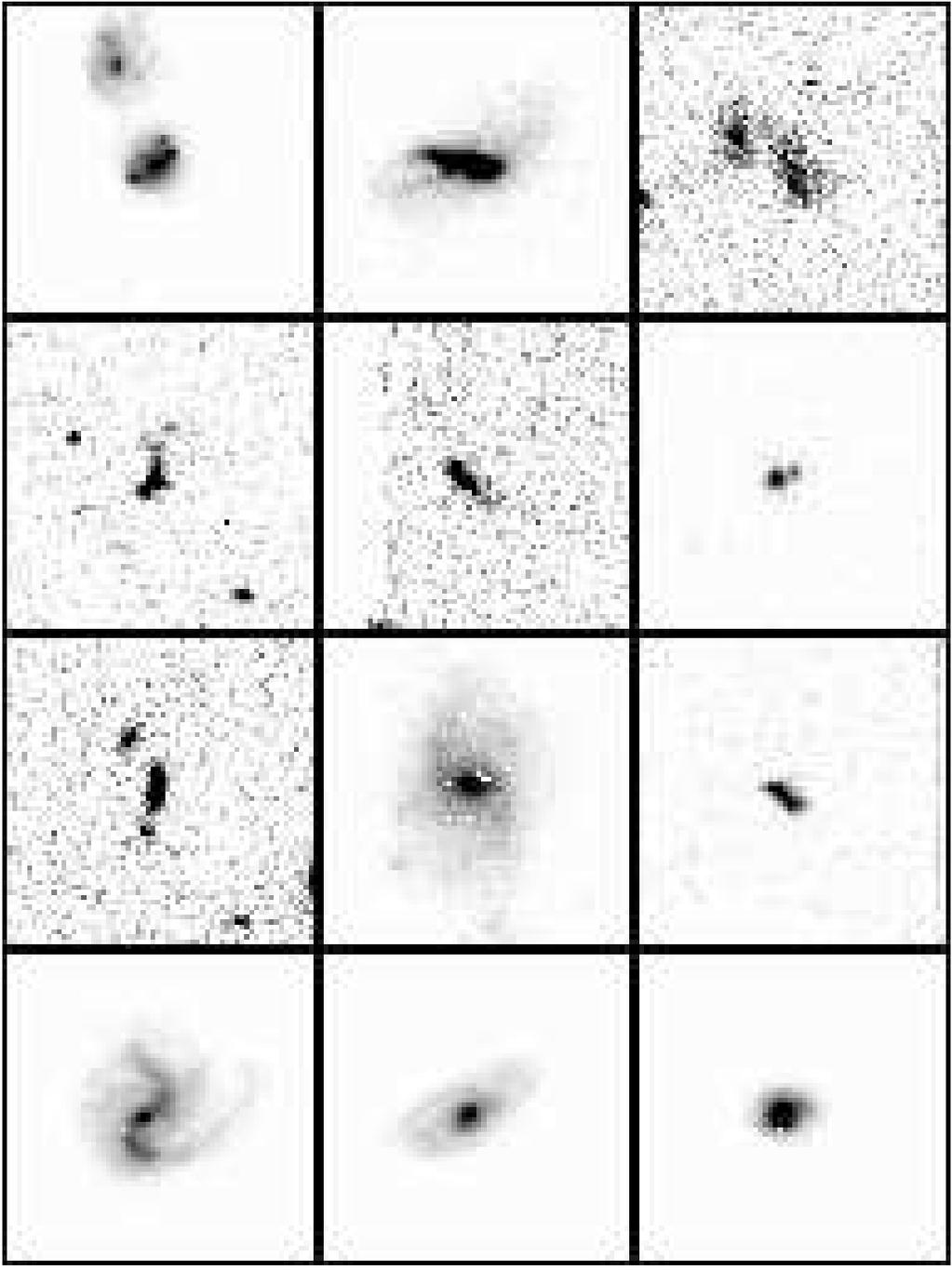
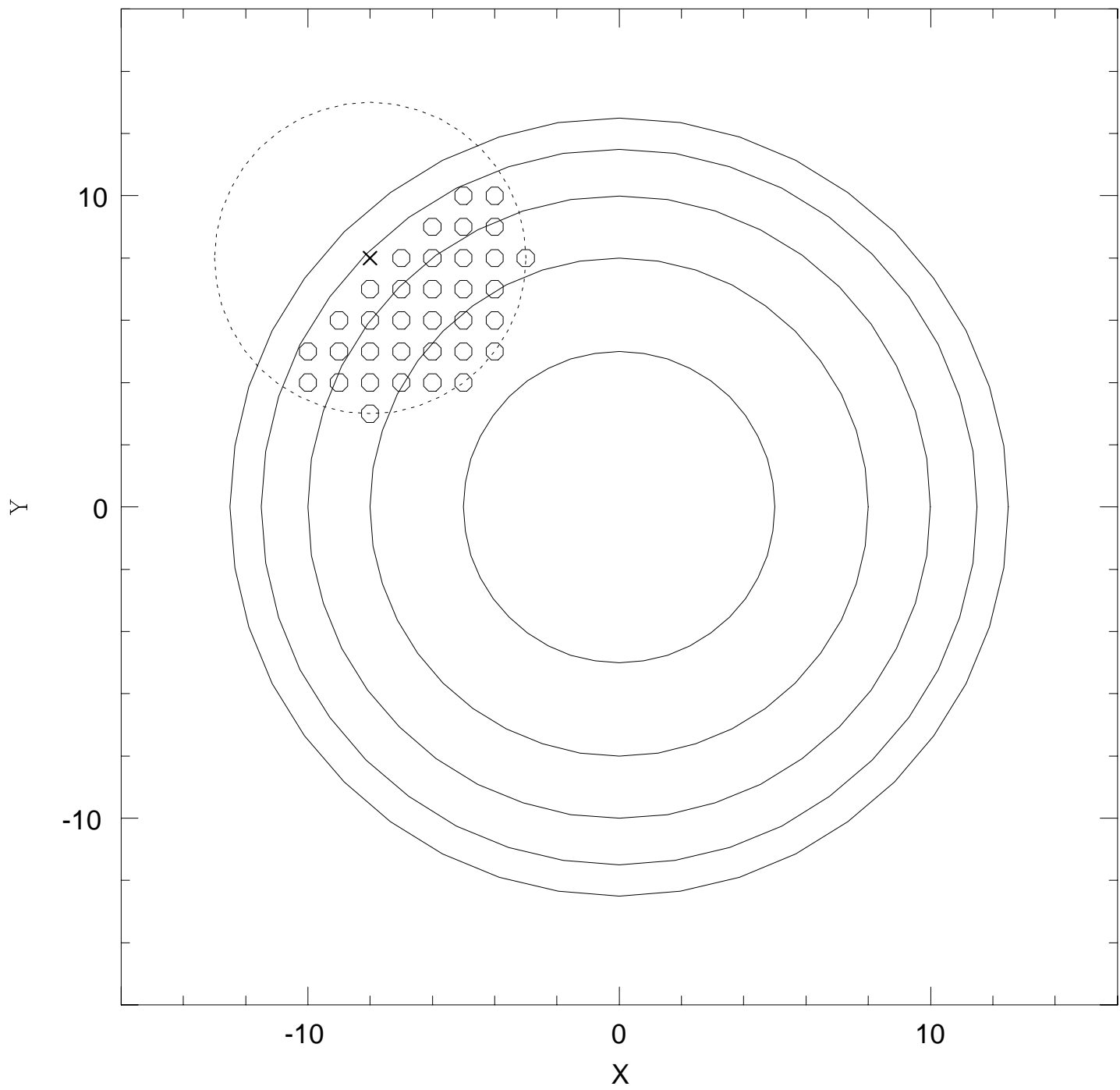
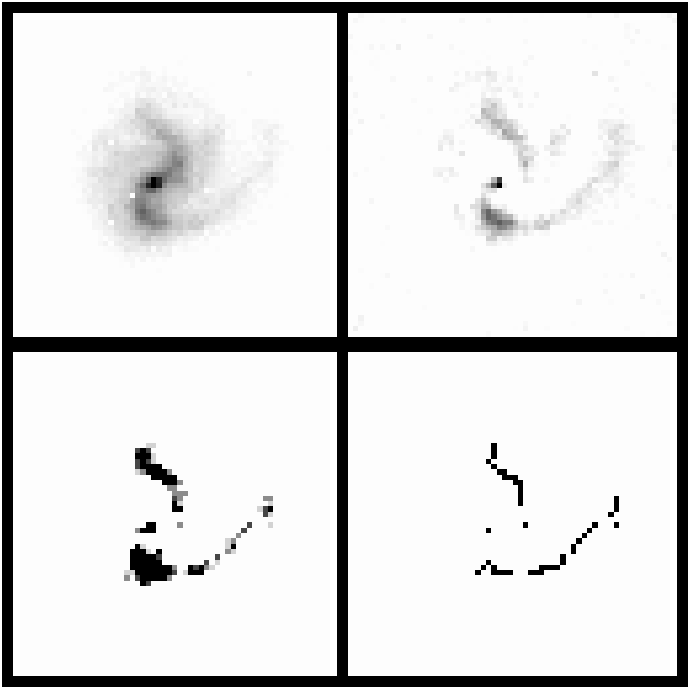
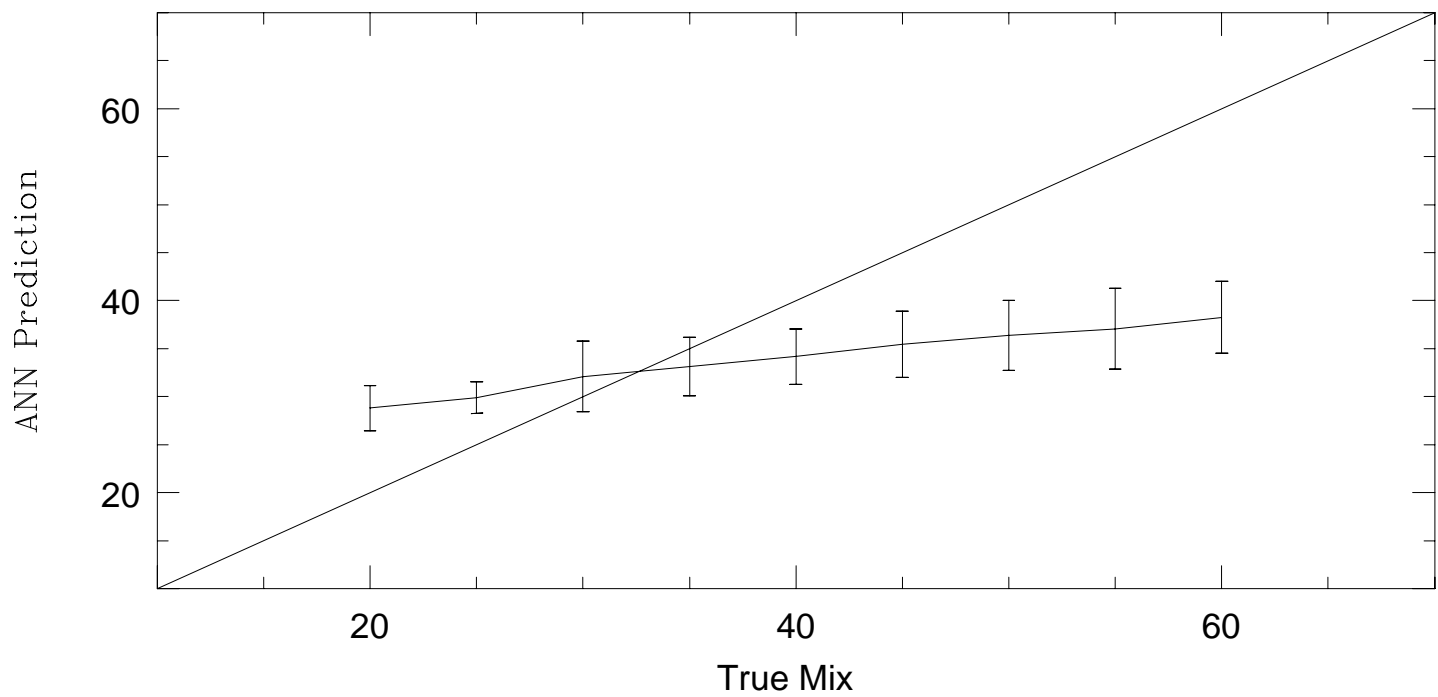
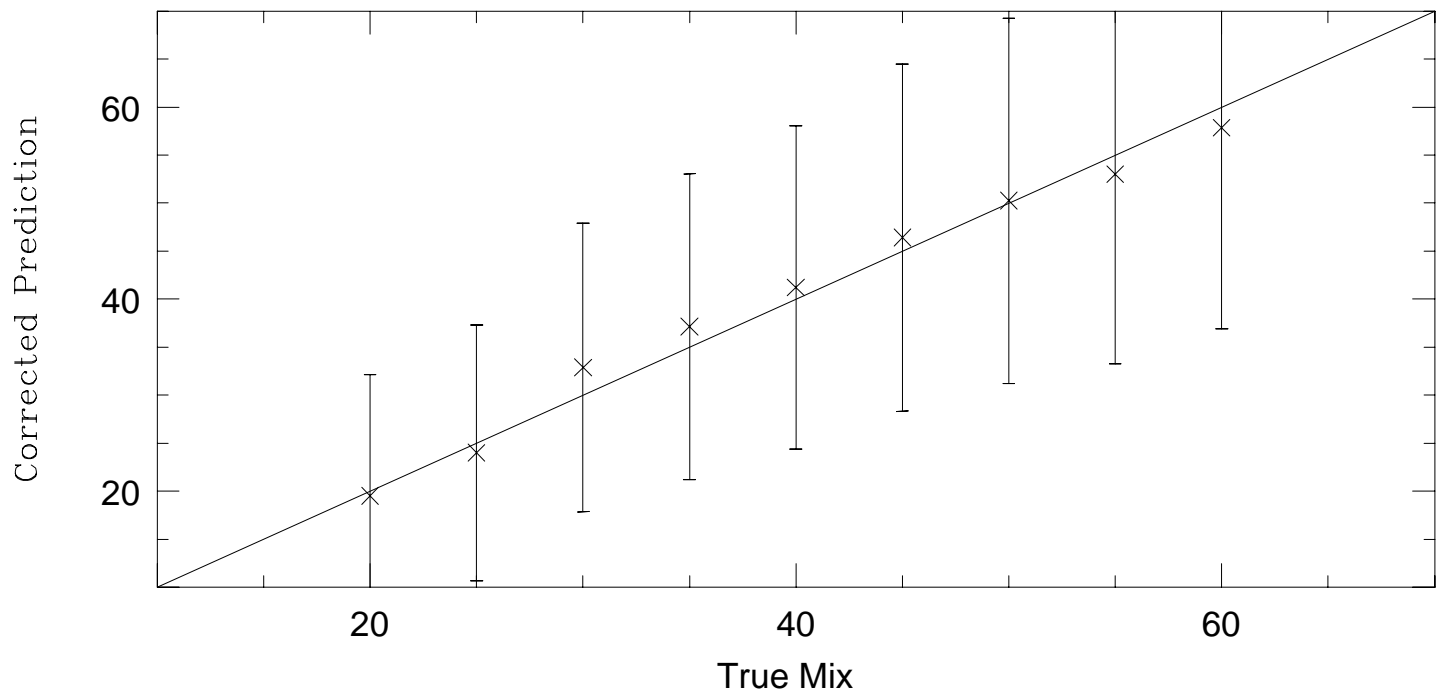
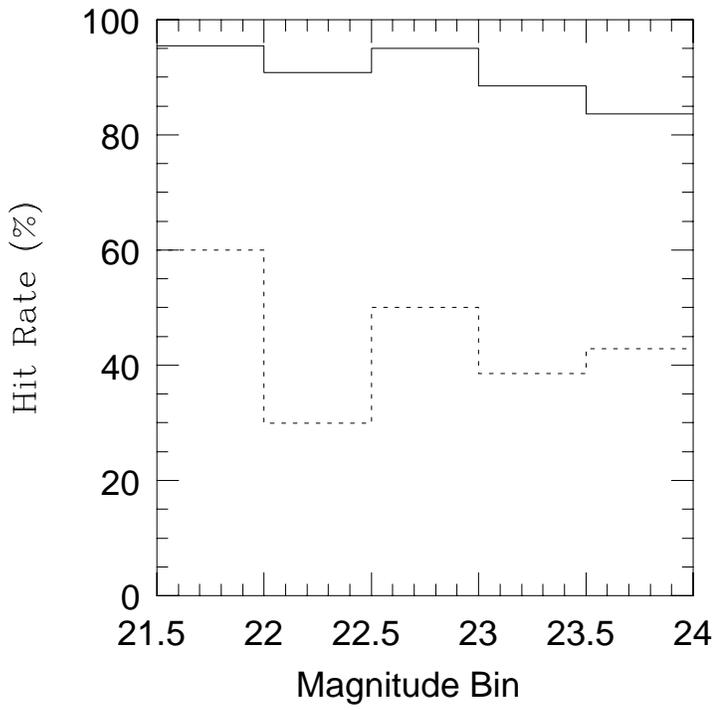
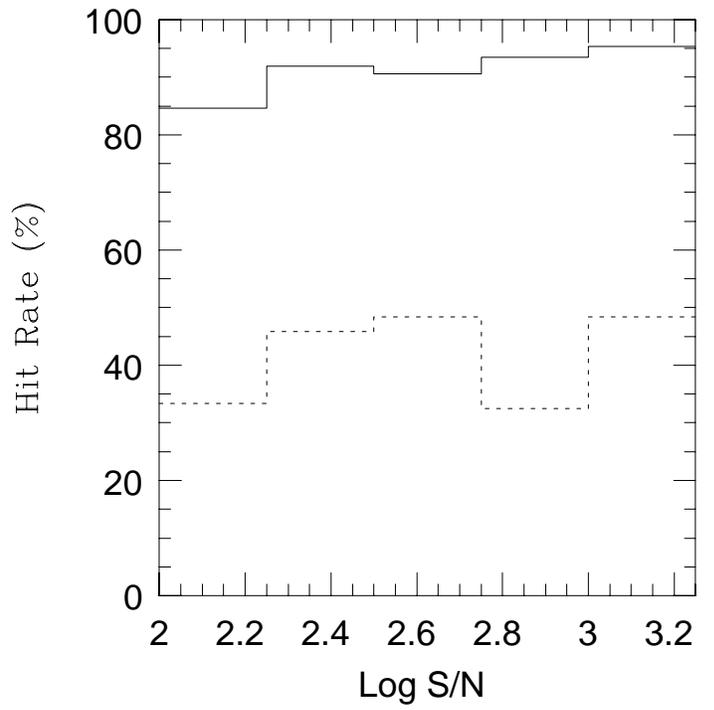
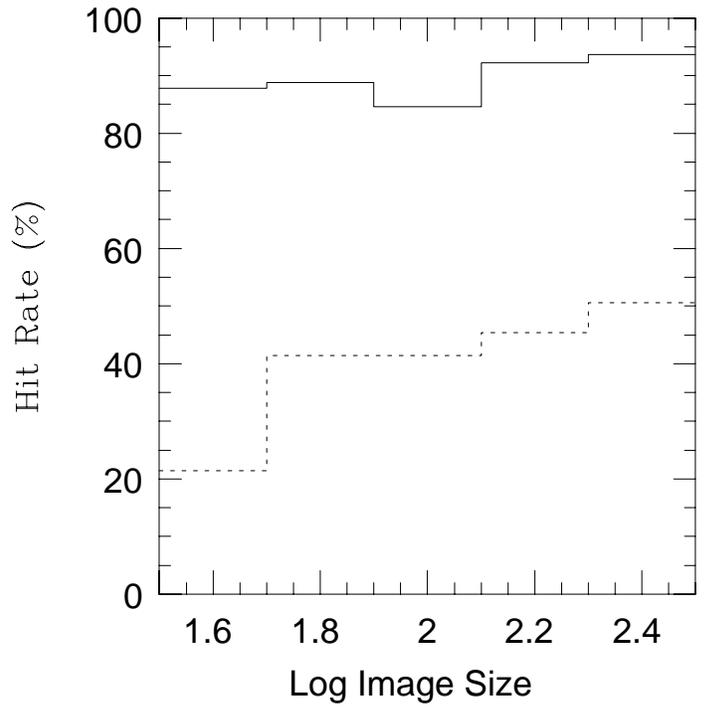
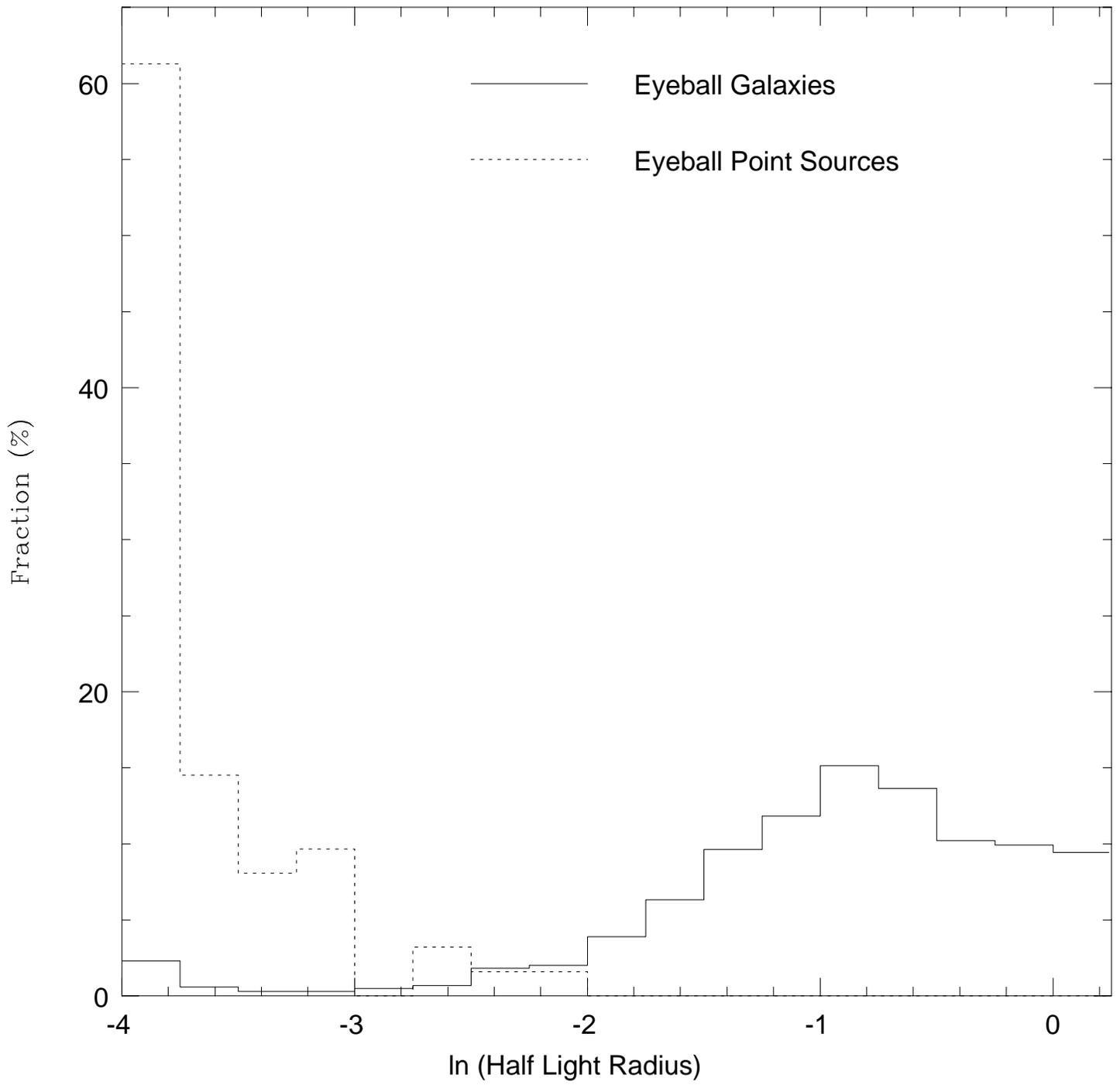
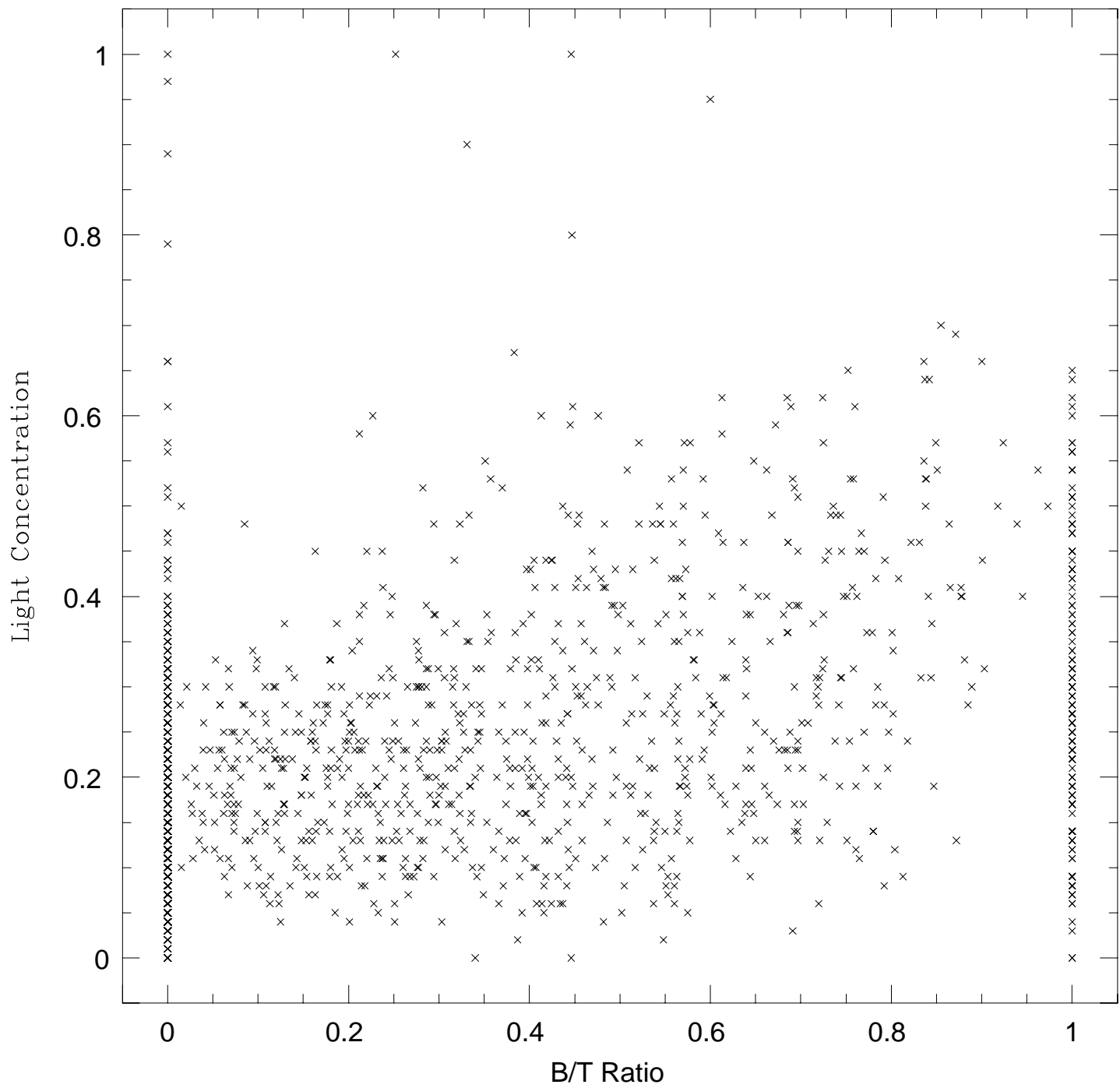
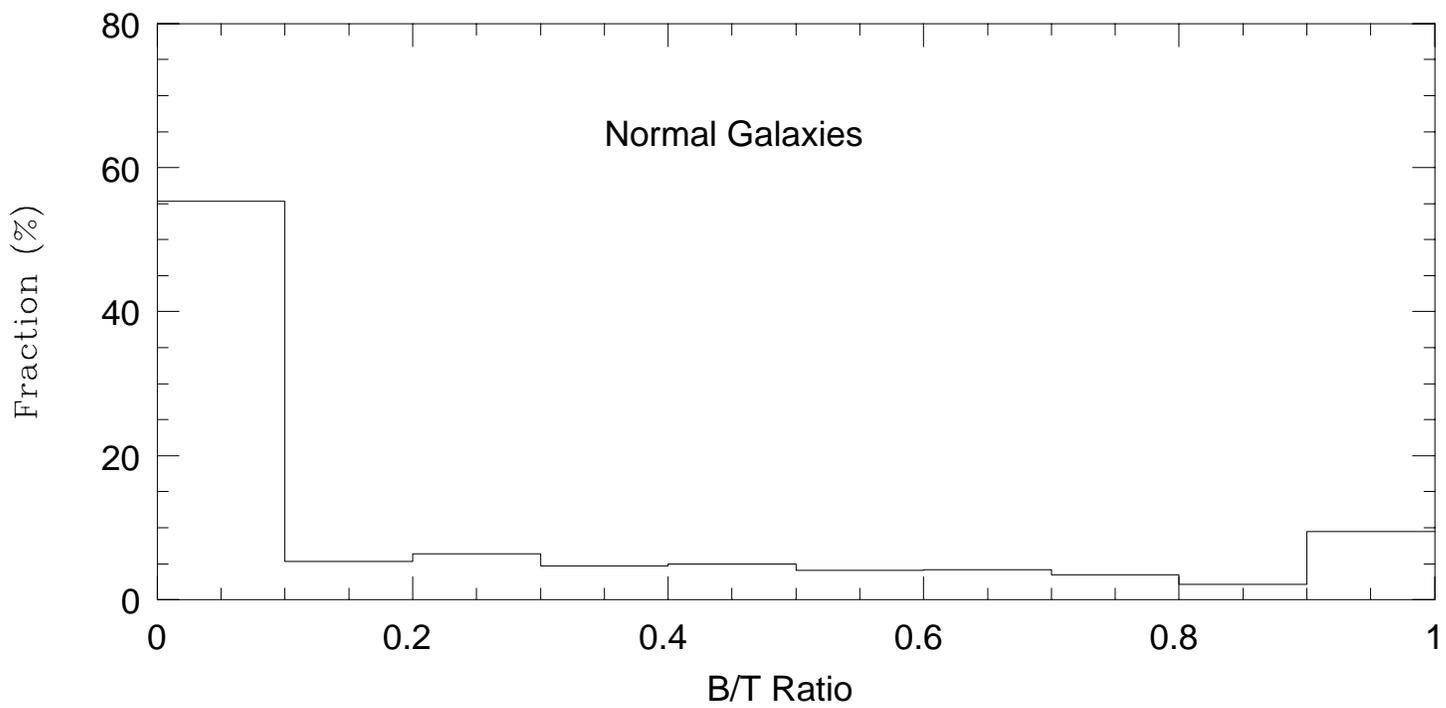
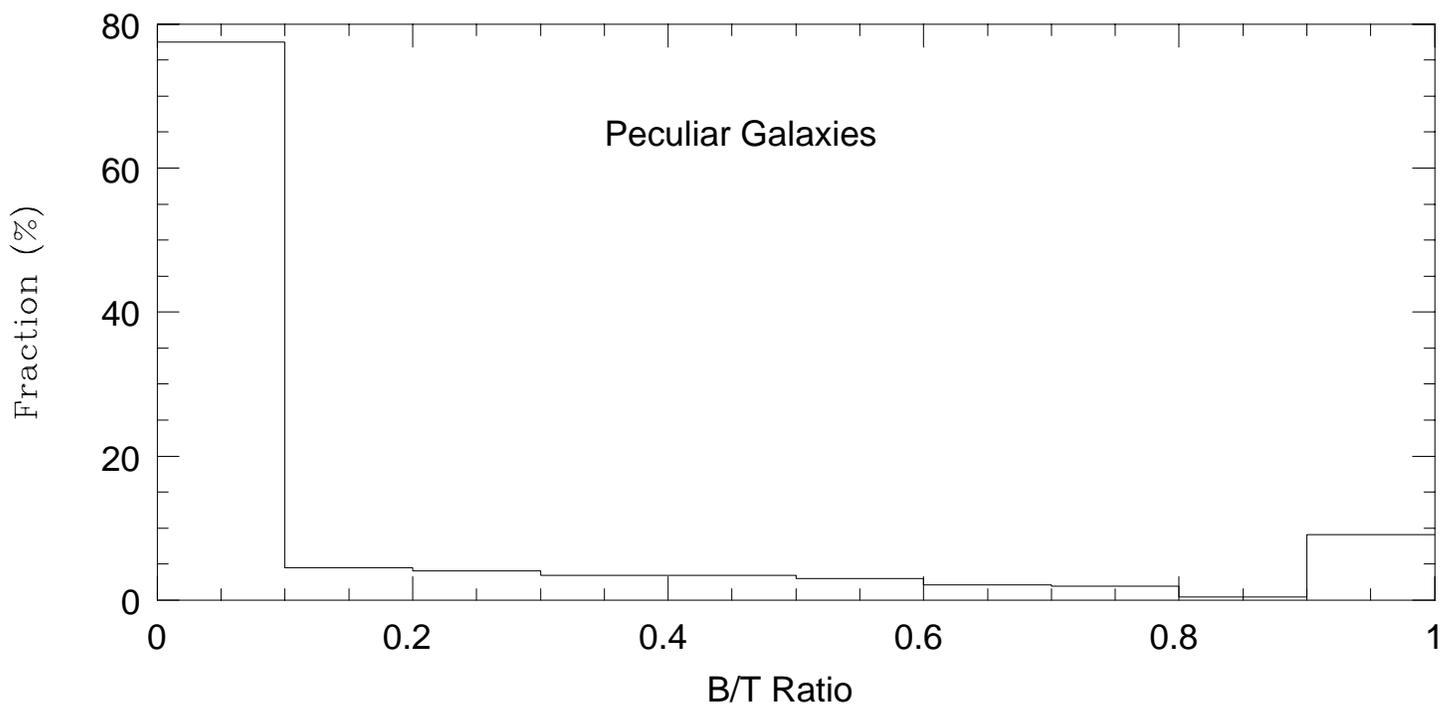
\end{document}